\documentclass[floats,superscriptaddress,pre]{revtex4}

\expandafter\let\csname equation*\endcsname\relax

\expandafter\let\csname endequation*\endcsname\relax

\usepackage{amsmath}

\usepackage{graphicx,psfrag,bbm,latexsym,color,dcolumn,bm,dsfont,bbm,
color,mathrsfs,bbold,latexsym,amsfonts,amssymb}

\usepackage{hyperref} 

\usepackage{amssymb}
\definecolor{myred}{RGB}{168,5,14}

\definecolor{editorcolor}{RGB}{0,0,0}

\begin{document}

\title
    {Spectrum of the tight-binding model on Cayley Trees and comparison with Bethe Lattices}

\author{M. Ostilli}
\affiliation{Instituto de F\'isica,  Universidade Federal da Bahia, Salvador--BA, 40170-115, Brazil}
    
\author{Claudionor G. Bezerra}
\affiliation{Departmento de F\'isica, Universidade Federal do Rio Grande do Norte, Natal--RN, 59078-970, Brazil}

\author{G. M. Viswanathan}
\affiliation{Departmento de F\'isica, Universidade Federal do Rio Grande do Norte, Natal--RN, 59078-970, Brazil}
\affiliation{National Institute of Science and Technology
                   of Complex Systems,
                   Universidade Federal do Rio Grande do Norte,
                   Natal--RN, 59078-970, Brazil}


\begin{abstract}
  There are few exactly solvable lattice models and even fewer solvable
  quantum lattice models. Here we address the problem of finding the 
  spectrum of the tight-binding model (equivalently, the spectrum of the adjacency matrix) on Cayley trees.  Recent
  approaches to the problem have relied on the similarity between
  Cayley tree and the Bethe lattice.  Here, we avoid to make any ansatz related to the Bethe lattice due to
  fundamental differences between the two lattices that persist even
  when taking the thermodynamic limit.  Instead, we show that one can
  use a recursive procedure that starts from the boundary and then use
  the canonical basis to derive the complete spectrum of the
  tight-binding model on Cayley Trees.
Our resulting algorithm is extremely efficient, as witnessed with remarkable large trees having hundred of shells.
We also shows that, in the thermodynamic limit, the density of states  
is dramatically different from that of the Bethe lattice.
\end{abstract}

\maketitle

\email{massimo.ostilli@gmail.com}

\section{Introduction}
Cayley trees, i.e., regular finite trees, provide the graph structure
of {lattice} models where exact {solutions} can be
derived.  This is the case for example of classical statistical
mechanics where the Ising model, or any other spin model, can be
easily solved by a recursive approach that starts from the boundaries
of the tree {(the leaves)}~\cite{Ostilli}.  There is no reason why
  the same  approach cannot be used for quantum mechanical systems.
Indeed, here we show precisely how to derive the spectrum of the
tight-binding {(TB)} model defined over {Cayley trees by an efficient} recursive approach
that starts from the boundary.  {Existing previous analyses of this
model~\cite{Derrida,Yorikawa,Aryal} provide results which are 
derived by using ansatz inspired by
the solution over the Bethe lattice~\cite{Mahan}. The Bethe lattice,
however, by definition, is an infinite tree {where each node has the same number of neighbors,}
whereas the Cayley tree has a boundary {(where each node has just one neighbor)}
that contains a finite fraction of the total number of nodes.
As a consequence, there is no
equivalence between {Cayley trees and Bethe lattices,
not even in the thermodynamic limit of the
former; we stress that interpreting the Bethe lattice as the bulk of a sufficiently large Cayley tree
is misleading due to the impossibility to remove the effects of the boundaries: no matter how
far we are from the boundaries, the local proprieties of the system will always depend on the chosen boundary conditions.
These issues have been extensively clarified in Ref.~\cite{Ostilli} where it was also pointed out that, especially in the older literature,
a lack of consensus on the nomenclature ``Bethe lattice'' versus ``Cayley tree'', generated further confusion}. 
  We note in particular that, despite the
  title ``Anderson model on a Cayley tree: the density of states'', {as we prove,} Ref.~\cite{Derrida} provides a result
  which holds for the Bethe lattice, not for the Cayley tree (see detailed discussion in Sec.~\ref{BBB}).
  {Defining the Bethe lattice as a tree where each node has the same number of neighbors
    (which is the original definition after Bethe~\cite{Bethe})
    might sound unsatisfactory from a physical and also numerical point of view since such a tree is necessarily infinite.
    However, one can smoothly approach the Bethe lattice as the
  thermodynamic limit
  of regular random graphs~\cite{Bollobas,Wormald},
  \textit{i.e.}, random graphs where each node has the same number of neighbors (as in the original definition of the Bethe lattice), and
  where loops (closed paths of links), in the thermodynamic limit, ``become negligible'' (there are only long loops, \textit{i.e.},
  loops whose minimal length
  grows with the logarithm of the number of nodes so that,
  locally, the graph looks like a tree). Yet, we warn again that the latter
  feature of the regular random graph does not allow us to identify the Bethe lattice as the bulk of a Cayley tree~\cite{Ostilli}.
  Observe that also in the regular random graph there is no boundary.
  Note in particular that, no matter how large is the regular random graph, its few loops are responsible for the existence of phase
  transitions: in general, loops cannot be simply thrown away. Such an issue becomes especially hard in frustrated systems,
  however, in the thermodynamic limit,
  the local properties of non-frustrated systems built on regular random graphs coincide with those of the Bethe lattice strictly
  defined as an infinite tree and for which there exists abundant literature~\cite{Zecchina,Parisi,Goltsev}, also in the
  quantum case~\cite{Derrida,Mahan,Biroli,Biroli2,Monthus}~\footnote{We stress that also in Ref.~\cite{Monthus} the name ``Cayley tree''
  is used in the place of ``Bethe lattice''; the same nomenclature issue applies often also in papers
  within the mathematical community~\cite{Utkir}.}}
  By contrast, papers that study the TB model on actual Cayley trees, or more general intrinsically finite trees,
  seem to be rarer and focused on specific problems such as, quantum walks~\cite{Jackson}, antiferromagnetism~\cite{Changlani}, and 
  Anderson's localization~\cite{Tikhonov,Sonner}. {See also the recent Ref.~\cite{Biroli2} for the different scenarios that emerge
  within many-body-localization when Bethe lattices and Cayley trees are compared}.
  On the other hand, the interest toward finite trees goes beyond a mere
  theoretical field. In fact, experimental implementations of Hamiltonians built on
  Cayley trees have been realized very recently for possible quantum simulators using Rydberg atoms~\cite{Simulation}.

  The above considerations call for a critical analysis of the issue Cayley tree \textit{vs} Bethe lattice
  {also within quantum mechanics. However, in the classical case, solving a system built on the top of a Cayley tree
  is quite trivial~\cite{Ostilli}, but one cannot say the same for the quantum case}.
  Moreover, we note that Ref. \cite{Yorikawa} considers general
  Cayley trees but its approach is
  essentially empirical,
  while Ref. \cite{Aryal} solves analytically the spectrum for Cayley trees with coordination number $z=3$
  by making use of a special basis but 
  it is
  difficult to figure out how to
  generalize the latter to arbitrary values of $z$, especially when $z$ is even.}
On the other hand, the {spectrum} of a model defined over a Cayley
tree should be derived in close analogy to the classical
case {by using the canonical basis, i.e., the natural basis represented by the nodes.}
The main aim of this work is to fill this gap between the
classical and quantum cases, showing in particular that there is no
need to make any ansatz related to the Bethe lattice.
{{In fact}, our analysis shows that the density of states $\rho(E)$ of the two models
  are dramatically different: in the thermodynamic limit of the Cayley tree {$\rho(E)$ tends to a collection of Dirac' s delta, symmetrically distributed around the central (and largest) one at $E=0$}, 
  whereas in the Bethe lattice $\rho(E)$ is a smooth bounded function of $E$~\cite{Derrida},\cite{Mahan}.
  Within our algebraic approach, the reason for such a result, consistent with~\cite{Yorikawa,Aryal}, emerges algebraically
  as a consequence
of the existence of a boundary --- in the Cayley tree --- that grows exponentially with the number of shells.}
{Our 
  algorithm is {extremely} efficient and can be applied to trees of any degree and size,
  as we show with remarkably large samples.

We stress that, despite an abundant literature about Cayley trees within classical physics, a careful analysis shows
that the spectrum of the TB model on Cayley trees was analyzed systematically in a few cases. 
In fact, concerning
the spectrum, an important distinction is in order among different models. Depending on the physical framework one is more
interested in, such as, electronic bands, quantum walks, relaxation dynamics, or mean first-passage times, three different
operators on graphs can be considered: the adjacency matrix $A$, the Laplacian $L=D-A$, 
and the normalized Laplacian $L'=I-D^{-1/2}AD^{-1/2}$, $D$ being the diagonal matrix, whose elements are the degrees
of the nodes of the graph (which are
the sums of the rows of $A$). On Cayley trees,
the spectrum of $A$ was analyzed in the above discussed Refs.~\cite{Yorikawa,Aryal},
the spectrum of $L$ was analyzed in Ref.~\cite{Lapl_CT}, while
the spectrum of $L'$ in Ref.~\cite{Zhang}. We mention also Refs. \cite{Kadanoff,Vicsek,Vicsek1,Zhang0,Zhang1,Zhang2,Zhang3}
for the study of the Laplacian
spectra over self-similar trees (note however that Cayley trees are not self-similar and therefore cannot be solved
by the spectral decimation approach~\cite{Kadanoff}).
It is known that the spectra and the physical properties of $A$, $L$, or $L'$,
{apart from the Bethe lattice case (for which $D$ is a trivial constant matrix),}
are dramatically different~\cite{Chen,Zhang,Lapl_vs_Adj}.
At any rate, within condensed matter physics,
the TB model {on Cayley trees}, which is the subject of this work, is the most appropriate one and amounts
to the adjacency matrix case {$A$}.

\section{Cayley Trees and labeling}
A Cayley tree of coordination number $z\geq 3$, is a simple finite graph defined as follows: Given a
root vertex labeled as 0, we attach to it $z$ neighbors whose set
constitutes the first shell of the graph. Then, each of these $z$
neighbors is attached to $z-1$ further neighbors ($b=z-1$ is also called the branching number). This completes the
second shell. The process keeps on until the desired last $L$-th shell
is formed. The resulting graph is a tree where each node has degree
$z$, except for the boundary nodes of the last $L$-th shell which,
by construction, have degree 1.  It is easy to see that the number of
nodes of the $k$-th shell, {$N_k$, and the total
number of nodes of the graph, $N$, are respectively given by
  \begin{eqnarray}
    \label{NL}
N_k=z(z-1)^{k-1}~,
  \end{eqnarray}}
  \begin{eqnarray}
    \label{N}
    N=1+\sum_{k=1}^{L} z(z-1)^{k-1}=\frac{z}{z-2}(z-1)^L-\frac{2}{z-2}.
  \end{eqnarray}
  
In Fig. \ref{fig1} we show an example of a Cayley tree of coordination number $z=3$ (or branching number $b=z-1=2$) with $L=3$ shells. 
Each node can be reached from the root vertex 0 by a unique {path of links, equivalent to a sequence of nodes}.
Therefore, we can conveniently attribute a label to each node
by using the {sequence} that leads to it, as explained in
Fig. \ref{fig1} with a little convention. This labeling is an important ingredient for the solution of the model.

  \begin{figure}[h]
  \centering
    \includegraphics[width=0.4\columnwidth,clip]{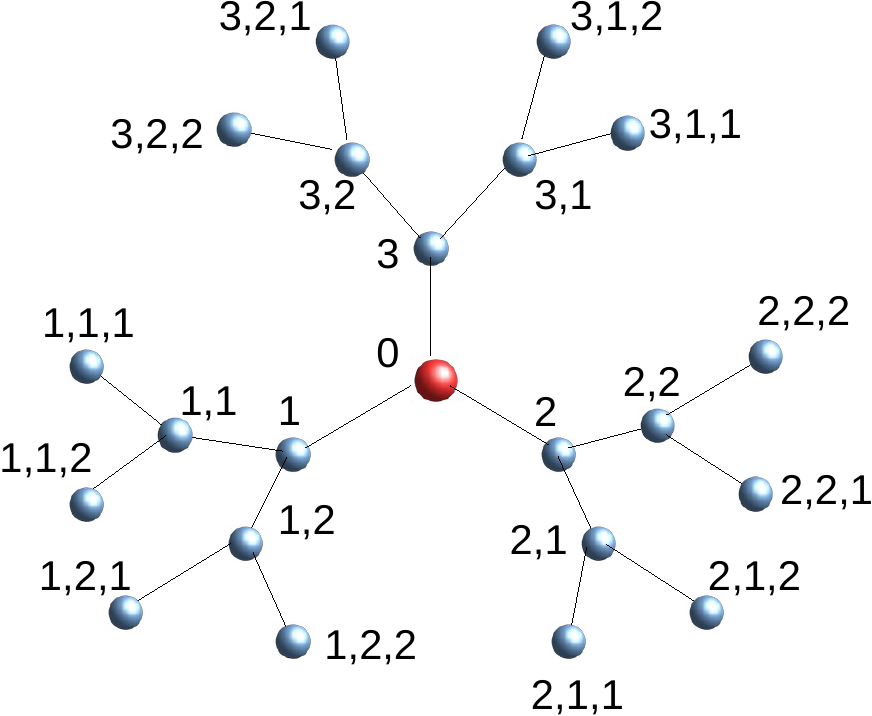}
  \caption{ 
    A Cayley tree of coordination number $z=3$ and $L=3$ shells. Each node can be reached from the root vertex 0 (the red central node) by a unique random walk that first selects
    the branch, 1, 2, or 3, then, at each successive jump, selects a node with sublabel 1 or 2 representing the position on the right and on the left, respectively, as seen by an observer standing
    at the center of the Cayley tree in a vertical direction with respect to the plane of the Cayley tree. The names of the nodes are then formed accordingly to the random walk sequences
    as shown in this example.
  }
  \label{fig1}
\end{figure}

{For later use, we introduce the following interesting identity
  \begin{eqnarray}
    \label{Np}
    N=(z-1)^L+2\sum_{k=0}^{L-1} (z-1)^{k}. 
  \end{eqnarray}
Equation (\ref{Np}) will be exploited in Sec. IV within the discussion of the degeneracy of the energy levels.} 
  
  \section{The tight-binding model}

 In condensed matter physics, the {TB} model
  has found successful application during several decades as a
  convenient and transparent model for the description of electronic
  structure in molecules and solids. In its relatively simple
  description of the electronic structure of a physical system, TB
  considers a solid described as a lattice formed by independent sites
  separated by the lattice constant. Each site is assigned a single
  atomic orbital, so that the electronic wave function may be written
  as a combination of localized atomic orbitals. Despite its
  simplicity, the precision of the TB approximation can be
  surprisingly good. For a given specific material the precision
  depends on the length of the independent wavefunctions compared to
  the lattice constant, i.e., it depends on the overlap between the
  wave functions of neighboring atoms. If the overlap is small
  enough, the approximation can be quite accurate, providing results
  in good agreement with more realistic models or even with
  experimental results. More sophisticated methods for electronic
  structure also make use of TB approach. For example, the eigenfunctions
  of the system can be written as a Linear Combination of Atomic
  Orbitals (LCAO), similar to the molecular orbital approach. Even
  though TB can be applied to non-crystalline solids, its most common
  application is in systems with translational symmetry, where the
  atomic orbitals satisfy Bloch’s theorem {(recent examples are
  graphene as well as boron-nitride sheets and nanotubes; for a review about TB model
  see for example Ref.~\cite{Economou}). It is important to note, however, that Cayley trees, as well as Bethe lattices,
  do not own (discrete) translational symmetry. In contrast, only Bethe lattices are self-similar, thanks to the lack of a boundary
  (and of a center) \cite{Ostilli}. These observations make it evident that the ``band-structures''
  on Cayley trees and Bethe lattices might be, not only not equivalent to those of finite dimensional lattices, but also  
  radically different between them. We shall return on this point later.}

Let us consider a single particle living on the nodes of a Cayley
tree. Each node is assigned a single atomic orbital. Thus, the set of
$N$ nodes, $0,1,\ldots, N-1 $, provides the canonical basis
$|0\rangle,|1\rangle,\ldots,|N-1\rangle $ defining a Hilbert space {of dimension $N$}
and, as mentioned  above, the electronic wave function may be written as a
linear combination of the localized atomic orbitals. We are interested
in the spectrum of the following tight-binding Hamiltonian with open
boundary conditions,
  \begin{eqnarray}
    \label{H}
    H=t\sum_{ < i,j > }{(|i \rangle \langle j|+|i \rangle \langle j|)}~,
  \end{eqnarray}
  where $t$ is the hopping constant, $i$ and $j$ are nodes of the Cayley tree,
  and $<i,j>$ {runs over non-ordered pairs with $i$ and $j$ nearest neighbors}.
{Observe that $H$ is nothing else but the adjacency matrix of the Cayley tree.}

Let $|\psi\rangle$ be an eigenvector of $H$ with eigenvalues $E$, $H|\psi\rangle=E|\psi\rangle$, and let $\psi(\cdot)=\langle \cdot |\psi\rangle$ be the components
of $|\psi\rangle$ on the canonical basis. 
From Eq. (\ref{H}) we have
\begin{eqnarray}
  \label{psi}
  t\sum_{j\in\mathcal{N}(i)} \psi(j)=E\psi(i)~,
\end{eqnarray}
where $\mathcal{N}(i)$ stands for the set of first neighbors of node $i$.
{In the following, we show how to derive the spectrum
by using a recursive procedure that starts from the boundary and use the canonical basis.}

\section{Main result: recursive equations and spectral properties for general $z$ and $L$}\label{Main}
Here we state our general result. Let be given a Cayley tree of coordination number $z$ and $L\geq 2$ shells
(if $L=1$, we have the eigenvalue $E=0$, which is $z-1$-th fold degenerate, and $E=\pm t\sqrt{z}$; see next section for details). 
Let us consider the following set of polynomials of the adimensional variable $\lambda=E/t$,
$P_0,P_1,\ldots,P_L$, built recursively by the following system 
\begin{eqnarray}
  \label{Recursive}
  \left\{
     \begin{array}{l}
       P_0(\lambda)=1,\\
       P_1(\lambda)=\lambda,\\
       P_{k+1}(\lambda)=\lambda P_k(\lambda)-{(z-1)}P_{k-1}(\lambda), \quad k=1,\ldots,L-1.
     \end{array}
     \right.
\end{eqnarray}
Each $P_k(\lambda)$ generated by the above system is a polynomial of degree $k$ in $\lambda$. Then, the complete
spectrum of the model~(\ref{H}) built
on the Cayley tree is provided by the roots of $P_1,\ldots,P_{L}$ and the roots of the equation $\lambda P_L=zP_{L-1}$. In other words,
for any $z$ and any $L\geq 2$, we have
\begin{align}
  \label{Spectrum}
  \mathrm{Spectrum}\left[\frac{H}{t}\right] = \cup_{k=1}^L\mathrm{~roots~of~} \left[P_k(\lambda)\right]~
  \cup \mathrm{~roots~of~}  \left[\lambda P_{L}(\lambda)-zP_{L-1}(\lambda)\right].
\end{align}

The spectrum is clearly symmetric around the solution $\lambda=0$ and, by direct inspection, it is possible to check
that {$\lambda\in[-2\sqrt{z-1},2\sqrt{z-1}]$}.
Moreover, our analysis {shows} that,
besides the geometric degeneracy of the eigenvalues, due to the above factorization,
there exists also an abundant algebraic degeneracy.
{In fact, the number of distinct eigenvalues is order $\mathop{O}(2(z-1)L)$
and, if we collect their absolute values in increasing order starting from $E=0$,
  it turns out that some eigenvalues have degeneracy $\mathrm{O}((z-1)^{L-k})$, for some $0\leq k\leq L$,
  while some others have degeneracy $\mathrm{O}(1)$, with the two groups alternated in a intricate manner, see
  for example Fig. 3. Note that the recursive procedure encoded in the spectral algorithm (\ref{Spectrum}) contains
  only the algebraic degeneracy. The analysis of the complete degeneracy of the single energy level
  is beyond the scope of the present work (and we are not aware of any closed formula able to provide such intricate degeneracies
  in general).   
  Nevertheless, we find three fundamental facts:\\
  i) The null eigenvalue has the following exact (and maximal) degeneracy 
\begin{eqnarray}
  \label{Deg0}
D(\lambda=0)=(z-1)^{L};
\end{eqnarray}
  ii) The maximal and minimal eigenvalues have no degeneracy 
\begin{eqnarray}
  \label{DegEmax}
D(\min \lambda)=D(\max \lambda)=1;
\end{eqnarray}
 iii) If we consider $2L+1$ bins covering symmetrically the whole interval $\in[-2\sqrt{z-1},2\sqrt{z-1}]$, so that each bin has
  width $4\sqrt{z-1}/(2L+1)$, and, from left to right, each bin has an associated index $k=-L,-L+1,\ldots,L-1,L$,
  the total degeneracy $D_{\mathrm{bin}}(k)$ within the bin with index $k$ turns out to be very well approximated by 
\begin{eqnarray}
  \label{Deg}
D_{\mathrm{bin}}(k)= (z-1)^{L-|k|}, \quad k=0,\pm 1,\ldots,\pm L.
\end{eqnarray}

The ansatz (\ref{Deg}) is justified by the fact that, besides interpolating Eqs. (\ref{Deg0})-(\ref{DegEmax}),
 due to Eq. (\ref{Np}), it also satisfies the correct normalization: $D_{\mathrm{bin}}(0)+2\sum_{k=1}^{L} D_{\mathrm{bin}}(k)=N$.}
  
  The eigenvalues associated to the last equation in (\ref{Spectrum}), \textit{i.e.},
  the roots of $\lambda P_{L}(\lambda)-zP_{L-1}(\lambda)$,
  deserve a special attention. According to Eq. (\ref{Deg}),  
  these are the least degenerate eigenvalues. In fact, the ground state belongs to this group and in the thermodynamic limit
  tends to $E=-2t\sqrt{z-1}$. It turns out that these eigenvalues, in general,
  are associated to eigenstates that are symmetric among the nodes that belong to the same shell.
  In fact, the equation $\lambda P_{L}(\lambda)-zP_{L-1}(\lambda)=0$ can be directly and more easily derived by ignoring
  the non symmetric eigenstates, \textit{i.e.}, by imposing that
  the amplitudes of the eigenstates depend solely on the shell index to which the given node belongs (see Sec. \ref{SS}).
It turns out that the symmetric eigenstates are similar to the eigenstates of the Bethe lattice, and yet,
as we shall discuss in more detail later, their respective eigenvalues are in general different.
  Note also that each eigenvalue can
  be associated to both symmetric and non symmetric eigenstates, as occurs, for example, with the null eigenvalue $E=0$.

  The above recursive procedure allows to find numerically all the eigenvalues very efficiently.
  For example, focusing on the spectra associated to the symmetric eigenstates,
  we proceed as follows:
  select random trials $\lambda$ in the range {$[-2\sqrt{z-1},2\sqrt{z-1}]$}, evaluate
$P_{L-1}$ and $P_L$ from the recursion system~(\ref{Recursive}),
and plot $|\lambda P_L-{z}P_{L-1}|$ versus $\lambda$. The local minima of this plot provide the location of the eigenvalues.
It turns out that such a numerical procedure is extremely efficient.
In fact, it allows to evaluate the spectra of any Cayley tree with quite large values of $L$.
In Fig.~\ref{fig3} we show the complete spectrum of the case $z=3$ with $L=10$, while in Fig.~\ref{fig2} we show the spectrum of
the symmetric eigenstates with $L=500$ where a ``quasi energy band'' structure emerges.

{\section{Proof of Eqs. (\ref{Recursive})-(\ref{Spectrum})}\label{Proof}
  Here we prove Eqs. (\ref{Recursive})-(\ref{Spectrum}). As we shall see, the derivation will provide us also important
  insights about the structure of the eigenstates that will be summarized in Sec.~\ref{General}.
It is convenient to introduce the following labeling of the nodes as illustrated in Fig. \ref{fig1} :
The central vertex is indicated by $0$; the $z$ vertices on the first shell are indicated by the numbers $m_1\in\{1,\ldots,z\}$;
for $k\geq 2$, the vertices on the $k$-th shell are indicated by the $k$-dimensional row vectors $(m_1,m_2,\ldots,m_k)$, 
where $m_1\in\{1,\ldots,z\}$, while $m_2,\ldots,m_k\in\{1,\ldots,z-1\}$. By definition, a vertex on the $k$-th shell
that has associated the vector $(m_1,m_2,\ldots,m_{k-1},m_k)$, is ``children'' of the ``parent'' vertex on the $k-1$-th shell
with associated vector $(m_1,m_2,\ldots,m_{k-1})$, i.e., from the vertex $(m_1,m_2,\ldots,m_{k-1})$ there emerge $z-1$ links pointing to the
vertices $(m_1,m_2,\ldots,m_{k-1},m_k)$. By using these definitions, given a Cayley tree with $L$ shells, the eigenvalue equation (\ref{psi})
written explicitly 
reads (we make use of the adimensional energy $\lambda=E/t$)

\begin{align}
  \label{A1}
  \left\{
     \begin{array}{l}
       \lambda\psi(m_1,m_2,\ldots,m_L)=\psi(m_1,m_2,\ldots,m_{L-1}),\\
       \lambda\psi(m_1,m_2,\ldots,m_{L-1})=\psi(m_1,m_2,\ldots,m_{L-2})+\sum_{m_L}\psi(m_1,m_2,\ldots,m_L),\\
       \vdots\\
       \lambda\psi(m_1,m_2,\ldots,m_{L-p})=\psi(m_1,m_2,\ldots,m_{L-(p+1)}) \\
       ~~~~~~~~~~~~~~~~~~~~~~~~~+ \sum_{m_{L-(p-1)}}\psi(m_1,m_2,\ldots,m_{L-(p-1)}), \quad 1\leq p\leq L-2,\\
       \vdots\\
       \lambda\psi(m_1)=\psi(0)+\sum_{m_2}\psi(m_1,m_2),\\
       \lambda\psi(0)=\sum_{m_1}\psi(m_1).
     \end{array}
     \right.
\end{align}

Note that, in Eq. (\ref{A1}),
it is understood that, for any row of the system with index $1\leq k\leq L$,
there appear $N_k=z(z-1)^{k-1}$ similar equations, the left hand side of each equation involving
a different $(m_1,m_2\ldots,m_k)$.

We now observe that the first equation of (\ref{A1}), 
\begin{eqnarray}
  \label{A2}
  \lambda\psi(m_1,m_2,\ldots,m_{L})=\psi(m_1,m_2,\ldots,m_{L-1}),
\end{eqnarray}
implies that, if $\lambda\neq 0$, $\lambda\psi(m_1,m_2,\ldots,m_{L})$
does not depend on the variable $m_{L}$. As a consequence, on plugging this information into the second equation of the system
we get 
\begin{eqnarray}
  \label{A3}
       \left[\lambda^2-(z-1)\right]\psi(m_1,m_2,\ldots,m_{L-1})=\lambda\psi(m_1,m_2,\ldots,m_{L-2}).
\end{eqnarray}
In turn, the above equation 
implies that, if $\left[\lambda^2-(z-1)\right]\neq 0$, the product $\left[\lambda^2-(z-1)\right]\psi(m_1,m_2,\ldots,m_{L-1})$
does not depend on the variable $m_{L-1}$. As a consequence, on plugging this information into the third equation of the system
we get 
\begin{eqnarray}
  \label{A5}
  && \left\{\lambda\left[\lambda^2-(z-1)\right]-\lambda(z-1)\right\}\psi(m_1,m_2,\ldots,m_{L-2})=~~~~~~~~~\nonumber \\
  && {\left[\lambda^2-(z-1)\right]}\psi(m_1,m_2,\ldots,m_{L-3}),
\end{eqnarray}
and so on, till the last equation of the system, which provides the following special relation
\begin{eqnarray}
  \label{A6}
  \lambda\psi(0)=z\psi(m_1).
\end{eqnarray}

From Eqs. (\ref{A2})-(\ref{A5}) and their generalizations,
we see that the amplitude of the wave function $|\psi\rangle$ on any node is proportional to
the amplitude of the wave function on the corresponding parent node. In other words, for $0\leq k\leq L-1$ we have
\begin{eqnarray}
  \label{A7}
  A_k\psi(m_1,m_2,\ldots,m_{L-k})=B_k\psi(m_1,m_2,\ldots,m_{L-k-1}), 
\end{eqnarray}
where $A_k$ and $B_k$ are two coefficients whose structure can be understood from Eqs. (\ref{A2})-(\ref{A5}) and their generalization.
By direct inspection we see that in general we have
\begin{eqnarray}
  \label{A8}
  A_k=P_{k+1}(\lambda), \quad B_k=P_k(\lambda), \quad 0\leq k\leq L-1,
\end{eqnarray}
where $P_k(\lambda)$ is a polynomial in $\lambda$ of degree $k$ which can be built recursively by the equation
\begin{eqnarray}
  \label{A9}
  P_{k+1}(\lambda)=\lambda P_k(\lambda)-{(z-1)}P_{k-1}(\lambda), \quad k=0,\ldots,L-1,
\end{eqnarray}
with the initial conditions $P_0\equiv 1$ and $P_1 \equiv \lambda$. Finally, on plugging Eqs. (\ref{A7})-(\ref{A8}) with $k=L-1$ 
into the last equation of the system, Eq. (\ref{A6}),
we get the equation for the eigenvalue $\lambda$:
\begin{eqnarray}
  \label{A10}
  \lambda P_{L}(\lambda)=z \lambda P_{L-1}(\lambda).
\end{eqnarray}

Note that we have reached the above result under the condition that $\lambda\neq 0$, $\lambda^2-(z-1)\neq 0$,
$\lambda\left[\lambda^2-(z-1)\right]-\lambda(z-1)\neq 0$,
and so on, which amounts
to say that $P_1(\lambda)\neq 0$, $P_2(\lambda)\neq 0$, $P_3(\lambda)\neq 0$, and so on.
On the other hand, if $\lambda$ is the root of one of these
polynomials, we can prove that $\lambda$ is also an eigenvalue of the system (\ref{A1}) as follows.
Let $\lambda$ be a root of $P_k(\lambda)$. It is convenient to distinguish the cases $k=1$ and $k\geq 2$.

If $P_1(\lambda)=0$, i.e., if $\lambda=0$, from the first and second equations of the system (\ref{A1}) we see
that we can always find an eigenvector $|\psi\rangle$, i.e., a non-null vector, such that
\begin{eqnarray}
  \label{A11}
  \left\{
  \begin{array}{l}
    \sum_{m_L}\psi(m_1,m_2,\ldots,m_L)=0, \quad \forall m_1,m_2,\ldots,m_{L-1}, \\
    \psi(m_1,m_2,\ldots,m_{L-p})=0, \quad \forall p\geq 1, \forall m_1,m_2,\ldots,m_{L-p}.
  \end{array}
  \right.
\end{eqnarray}
Note that the equations $\sum_{m_L}\psi(m_1,m_2,\ldots,m_L)=0$ admit certainly non trivial solutions,
for we can always find amplitudes on the $L$-th shell
with suitable signs such their sum is 0.

Suppose now that $P_k(\lambda)=0$ for some $k$ with $k\geq 2$. 
From Eqs. (\ref{A7})-(\ref{A8}) we have in particular the following two groups of relations
\begin{eqnarray}
  \label{A12}
  \left\{
  \begin{array}{l}
    P_{k}(\lambda)\psi(m_1,m_2,\ldots,m_{L-k+1})=P_{k-1}(\lambda)\psi(m_1,m_2,\ldots,m_{L-k}),\\
    P_{k+1}(\lambda)\psi(m_1,m_2,\ldots,m_{L-k})=P_{k}(\lambda)\psi(m_1,m_2,\ldots,m_{L-k-1}).
  \end{array}
  \right.
\end{eqnarray}
Due to the fact that $P_k(\lambda)=0$, we see
that we can always find an eigenvector $|\psi\rangle$, i.e., a non-null vector, such that
\begin{eqnarray}
  \label{A13}
  \left\{
  \begin{array}{l}
    \psi(m_1,m_2,\ldots,m_{L-k+1})\neq 0,  \quad \mathrm{for~some~} m_1,m_2,\ldots,m_{L-k-1}, \\
    \psi(m_1,m_2,\ldots,m_{L-p})=0, \quad \forall p\geq k, \forall m_1,m_2,\ldots,m_{L-p}.
  \end{array}
  \right.
\end{eqnarray}
{On plugging Eqs. (\ref{A13}) into the $k$-th row of the system (\ref{A1}) we are left with the condition
  $\sum_{m_{L-k+1}}\psi(m_1,m_2,\ldots,m_{L-k+1})=0$ which admits certainly non trivial solutions. 
  Once such components
$\psi(m_1,m_2,\ldots,m_{L-k+1})$ have been found,
the remaining of the equations of the system (\ref{A1}) are solved by iterating (\ref{A7}) backwards obtaining the following solution 
\begin{eqnarray}
  \label{A14}
  \left\{
  \begin{array}{l}
    \psi(m_1,m_2,\ldots,m_{L-k+q})=\frac{P_{k-q}(\lambda)}{P_{k-1}(\lambda)}\psi(m_1,m_2,\ldots,m_{L-k+1}),
    \quad 2 \leq q\leq k, \quad \forall m_1,m_2,\ldots,m_{L-k+q},\\
    \\
    \psi(m_1,m_2,\ldots,m_{L-p})=0, \quad \forall p\geq k, \quad \forall m_1,m_2,\ldots,m_{L-p}.
  \end{array}
  \right.
\end{eqnarray}
Since we have found an explicit solution of the eigenstate,
we have then proven that, if for some $k\in\{1,\ldots,L-1\}$ we have $P_k(\lambda)=0$, 
then, $\lambda$ is also an eigenvalue of the system (\ref{A1}).
} 

By summarizing, if an eigenvalue $\lambda$ is not a root of the polynomials $P_k(\lambda)$ for $k\in\{1,\ldots,L-1\}$,
then, $\lambda$ must be solution of Eq. (\ref{A10}); however, if $\lambda$ is a root of at least one of the polynomials
$P_k(\lambda)$, then, $\lambda$ is an eigenvalue (regardless whether $\lambda$ {also} is, or is not, solution of Eq. (\ref{A10})).
This fact completes the proof of
Eqs. (\ref{Recursive})-(\ref{Spectrum}).
}

  \begin{figure}[h]
  \centering
    \includegraphics[width=1.0\columnwidth,clip]{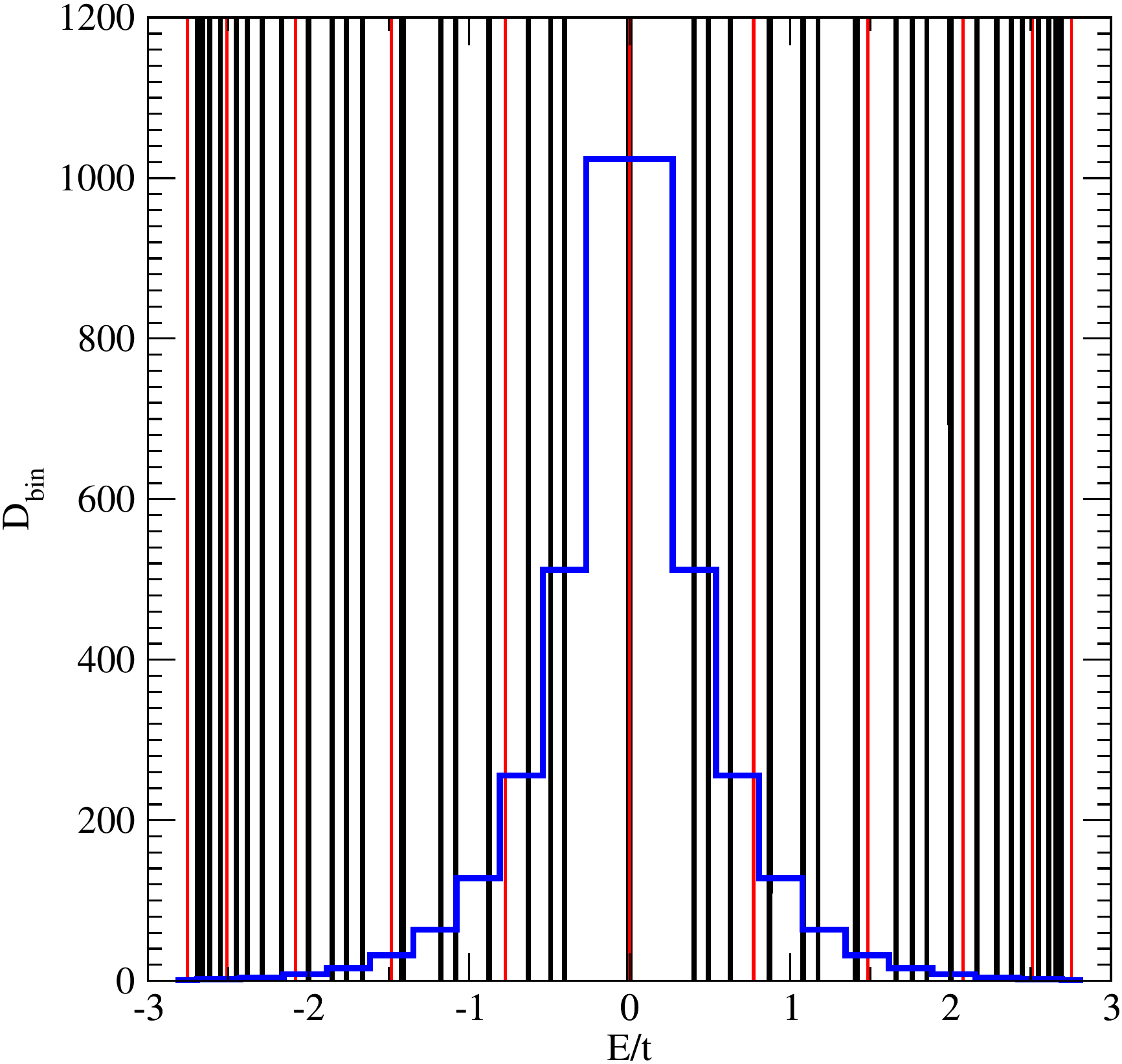}
  \caption{{
    Spectrum of a Cayley tree with {$z=3$ and $L=10$ shells}. The spectrum is obtained by solving Eqs.~(\ref{Recursive})-(\ref{Spectrum})
    via $10^5$ $\lambda$ trials in the range
    $[-2\sqrt{2},2\sqrt{2}]$. Black lines correspond to all possible eigenvalues while the red lines correspond
    to eigenvalues associated to symmetric eigenstates, which include the ground state.
    Note that some eigenvalues are associated to both symmetric and non symmetric
    eigenstates, as it occurs for the null eigenvalue. {The blue symmetric stepwise function is the plot of $D_{\mathrm{bin}}$,
    Eq. (\ref{Deg}), providing the total degeneracy of the eigenvalues that fall
    within each bin}.
  }}
  \label{fig3}
\end{figure}

{\section{Spectra of the symmetric states}\label{SS}}
As we have mentioned it before, some of the eigenstates, as e.g. the ground state, are symmetric among nodes that belong to the same shell.
The amplitudes of these eigenstates can therefore be indicated by using just the shell index to which the given node belongs. In other words we can identify the amplitudes of the symmetric eigenstates
as {($0-$th shell means central node)
 \begin{align}
     \label{Symm}
   \psi(i)\equiv \psi_k, \quad \forall ~i \in ~k-\mathrm{th}~\mathrm{shell}, ~k=0\ldots,L.
 \end{align}
 Eq. (\ref{Symm}) leads to a great simplification of the equations} for the eigenvalues associated to the symmetric eigenstates.
 {Let $\lambda=E/t$}.
 The eigenvalue problem that starts from the boundary shell $L$ can be written in the form
 \begin{align}
     \label{Symm1}
   A_k\psi_{L-k}=B_k\psi_{L-k-1}, \quad k=0,\ldots,L-1,
 \end{align}
where $A_k$ and $B_k$ are two suitable coefficients. {Then, as similarly done in Sec. \ref{Proof} for the general case, we get}
\begin{align}
     \label{Symm2}
   A_k=P_{k+1}, \quad B_k=P_{k}, 
 \end{align}
where $P_k$ is a polynomial of degree $k$ in $\lambda$ which is built recursively by
{the same general recursive system~\ref{Recursive}}.
Finally, once $P_{L-1}$ and $P_L$ have been built by using the recursion system,
the eigenvalue problem reduces to the following equation in $\lambda$
\begin{eqnarray}
  \label{Symm4}
{\lambda}P_L={z}P_{L-1}. 
\end{eqnarray}
The above procedure allows to find all the eigenvalues of the symmetric eigenstates. For example, it is easy to check that
${\lambda}=0$
is (also) a solution of the symmetric spectrum whenever $L$ is even.

\section{General properties of the eigenstates}\label{General}
The above analysis was mostly limited to the search for the eigenvalues of the model and it was based on a recursive procedure that starts from the boundary nodes. 
In order to get the eigenstates associated to a given eigenvalue $E$, one has to simply use the reversed path by starting from the central node 0. For example, {for $z=3$}, in the case of the ground state,
once its eigenvalue $E$ is plugged into the equation for {$\psi(1)$}, up to a normalization constant
this allows the determination of the amplitudes on the second shell, and so on, till reaching again the boundary nodes.
{See also Appendix for illustrative examples}.
By applying such a scheme and making use of the {insights gained in Sec. \ref{Proof}},
the following general features emerge:

\begin{itemize}
\item The ground state {(and therefore also the most exited state):} is unique ($D=1$); 
  {each node has a positive amplitude} (or, more precisely, there are no phases);
  it is symmetric over the different branches (i.e., if $i$ and $j$ belong to the same shell, then $\langle i|\psi\rangle =\langle j|\psi\rangle$);
  the maximal amplitude corresponds to the root vertex 0, while the amplitude decays monotonically for increasing shells (and it is therefore minimal on the boundary).

  \item All other symmetric eigenstates, i.e., those for which $\lambda=E/t$ is solution of the equation
    $\lambda P_L(\lambda)={z}P_{L-1}(\lambda)$, {have some shells with null amplitude or alternated signs}.  

  \item {As shown in the general proof in Sec. \ref{Proof},
    if $\lambda=E/t$ is root of some $P_k(\lambda)$, with $k\geq 1$, 
    then, according to Eqs. (\ref{A14}),
    the corresponding eigenvectors have non-null components only on the shells $L-k+1,L-k,\ldots,L$,
    with the components on the $L-k+1$ shell satisfying the condition $\sum_{m_{L-k+1}}\psi(m_1,m_2,\ldots,m_{L-k+1})=0$.
    Among the non trivial solutions of the latter, regardless of $z$,
    we can even build
    eigenvectors having just two non-null opposite components, all the others components being null. For example, for $z=3$ we can
    take $\psi(m_1,m_2,\ldots,m_{L-k},1)=1$, $\psi(m_1,m_2,\ldots,m_{L-k},2)=-1$,
    while $\psi(m^{'}_1,m^{'}_2,\ldots,m'_{L-k+1})=0$
    whenever there exists at least an index $i$ in the set $\{1,\ldots,L-k\}$ such that $m^{'}_i\neq m_i$. 
  }  
  
\item {The null eigenvalue $E=0$ is maximally degenerate with a degeneracy that grows exponentially with the number of shells.
  In fact, when $E=0$, the $N_L$ equations on the boundary shell determine
  that the amplitudes on the $(L-1)$-th shell must be null, which, in turn, determine that also
  the amplitudes on the $(L-2)$-th shell must be null, etc., till reaching a system involving only the $N_L$ amplitudes of the
  boundary shell
  together with the amplitudes of the first shell, for $L$ odd, or the amplitude of the central node, for $L$ even.
  As a consequence, with respect to the original eigenstate problem that, in general, for given $E$, involves
  $N$ equations with $N$ variables, 
  when $E=0$, we are left with a homogeneous linear system having $N_V\simeq N_L$ number of variables
  and $N_E\simeq N_{L-1}$ number of equations. Therefore, by using the explicit expression for $N_k$, Eq. (\ref{NL}), we see that
  such a system has $\infty^{D(E=0)}$ non trivial solutions, where $D(E=0)=N_V-N_E\sim (z-1)^L$. Direct inspection shows that it is exactly
  $D(E=0)=(z-1)^L$.
For $L$ even, there is also a special eigenstate
   which is symmetric among the branches (see {previous} section).
   Finally, in contrast to the symmetric eigenstate (which is present only for $L$ even), {within} the boundary
   shell, most of these eigenstates are formed by a combination of just a few nodes
   (in half cases just two) with same weights but alternated sign.    
  }
\end{itemize}  

  {\section{Comparison with the Bethe lattice}\label{BBB}}
On the Bethe lattice, all eigenvalues have the 
form $E=2\sqrt{z-1}\cos(\theta)$, with $\theta \in [0,2\pi)$~\cite{Mahan}.
  For $L\to \infty$, the spectrum of the symmetric eigenstates on the Cayley tree seems to tend to the spectrum
  of the symmetric eigenstates of the Bethe lattice~\cite{Mahan}. In particular,
direct inspection shows that, for $L\to\infty$, the ground state eigenvalue of the two cases coincide: $E=-2t\sqrt{z-1}$.
Furthermore, for finite $L$, most eigenvalues can be written in the Bethe lattice-like form as~\cite{Yorikawa}
\begin{eqnarray}
  \label{Bethe0}
  && E=2t\sqrt{z-1}\cos\left(\frac{{\pi} m}{l+1}\right), \quad 
  m=1,\ldots,l, \quad l=1,\ldots,L~.
 \end{eqnarray}
It is important to stress however that not all the eigenvalues of $H$ can be expressed as in (\ref{Bethe0}) via a rational fraction
of $\pi$.
A particularly important counter example concerns the ground state {energy}, for which $\theta$ is irrational
(see Ref. ~\cite{Aryal} for the case $z=2$).
More importantly, the similarity between the spectrum of Cayley tree and Bethe lattice breaks in some crucial points
even in the thermodynamic limit.
First of all, note that on the Bethe lattice the spectrum is continuous with no gaps. In fact,
whereas on the Bethe lattice for any $\theta \in [0,2\pi)$ there exists 
an eigenvalue $E=2\sqrt{z-1}\cos(\theta)$, 
  this is not guaranteed to be the case on the Cayley tree where it might occur that, even in the $L\to \infty$ limit, some $\theta$
  are missing, as e.g. Fig. \ref{fig2} leads to guess.
  Notice in particular that, as discussed in the previous section,
  within the symmetric spectrum, the eigenvalue $E=0$ is present only for $L$ even. This prevents
  therefore to form a true thermodynamic limit of the symmetric eigenstate associated to $E=0$.
  However, the most peculiar discrepancy between the Cayley tree and the Bethe lattice concerns
  the absence of the non symmetric states in the latter, which in turn implies a dramatic difference
  between their respective density of states. {Compare for example Figs. \ref{L_4} and \ref{DOS_BL}. Clearly, for the Cayley tree,
  for any $L$ finite we cannot use the concept of density of states. However, it is pretty clear that even in the thermodynamic limit,
  where a density of states $\rho(E)$ can be defined also for the Cayley tree, we have a dramatic difference.
  In fact, we have $\rho(E)\simeq \lim_{L\to\infty}\frac{\sum_{k}D_{\mathrm{bin}}(k)\chi_k(E) }{\sum_{k}D(E_k)}$, where $D_{\mathrm{bin}}(k)$
    is given by Eqs. (\ref{Deg}) and $\chi_k(E)$ is the characteristic function taking value 1 when $E$ falls within the bin with index $k$, and 0 otherwise. In particular, $\rho(E)$ turns out to be exponentially peaked in $E=0$. By contrast},  
  the density of states of the Bethe lattice is a bounded function of $E$~\cite{Mahan}: 
\begin{eqnarray}
  \label{rhoBethe}
  \rho_{BL}(E)=\frac{z}{2\pi}\frac{\sqrt{A^2-E^2}}{z^2-E^2},
\end{eqnarray}
where $A=2\sqrt{z-1}$.
Notice that, as we have anticipated in the introduction, also the density of states derived in Ref.~\cite{Derrida} coincides with
$\rho_{BL}(E)$~\footnote{In Ref.~\cite{Derrida}, the density of states is given in terms
  of an integrated density of states $n(\theta)$ written as a function of the angle $\theta=\cos^{-1}(E/A)$ (see therein Eq. (20)).
  On calculating $\partial/\partial\theta(1-n(\theta))\partial \theta/\partial E$, we have verified
  that we get exactly the density of states of the Bethe lattice $\rho_{BL}(E)$ as given by Eq. (\ref{rhoBethe}).}.
 
It might be questioned that the above density of states $\rho_{BL}(E)$ of the Bethe lattice derived in the Refs.~\cite{Derrida}
and ~\cite{Mahan}, might be incomplete because both derivations were
based on the symmetric ansatz imposed on the eigenstates~\footnote{Concerning Ref.~\cite{Derrida},
  a careful analysis of the paper shows that, while in their initial setting the authors consider general eigenstates for
  a general tree,
  the therein equations are then applied to a disordered model in which the symmetric ansatz is tacitly assumed, also when
  the absence of disorder is considered as a special case (see therein Eq. (20)).}.
  It remains open therefore the question whether the Bethe lattice admits also non symmetric
  eigenstates leading perhaps to the same density of states {of the Cayley tree}. However, this does not seem to be
  possible due to the intrinsic differences between the Cayley tree and the Bethe lattice, where only
  the former has a boundary. In fact, if we look back at the eigenstates associated to $E=0$, it is precisely thanks to the boundary
  having $N_L=z(z-1)^{L-1}$ terminal nodes, that there exist $(z-1)^L$ independent solutions, each having null amplitudes throughout the tree,
  except on the boundary and on the central node (for $L$ even), or on the boundary and on the first shell (for $L$ odd).
  As soon as we try to remove the boundary,
  the homogeneous system corresponding to $E=0$ admits only the symmetric solution, the exponential degeneracy disappears,
  and the density of states becomes the bounded
function $\rho_{BL}(E)$.

\begin{figure}[h]
  \centering
  \includegraphics[width=0.5\columnwidth,clip]{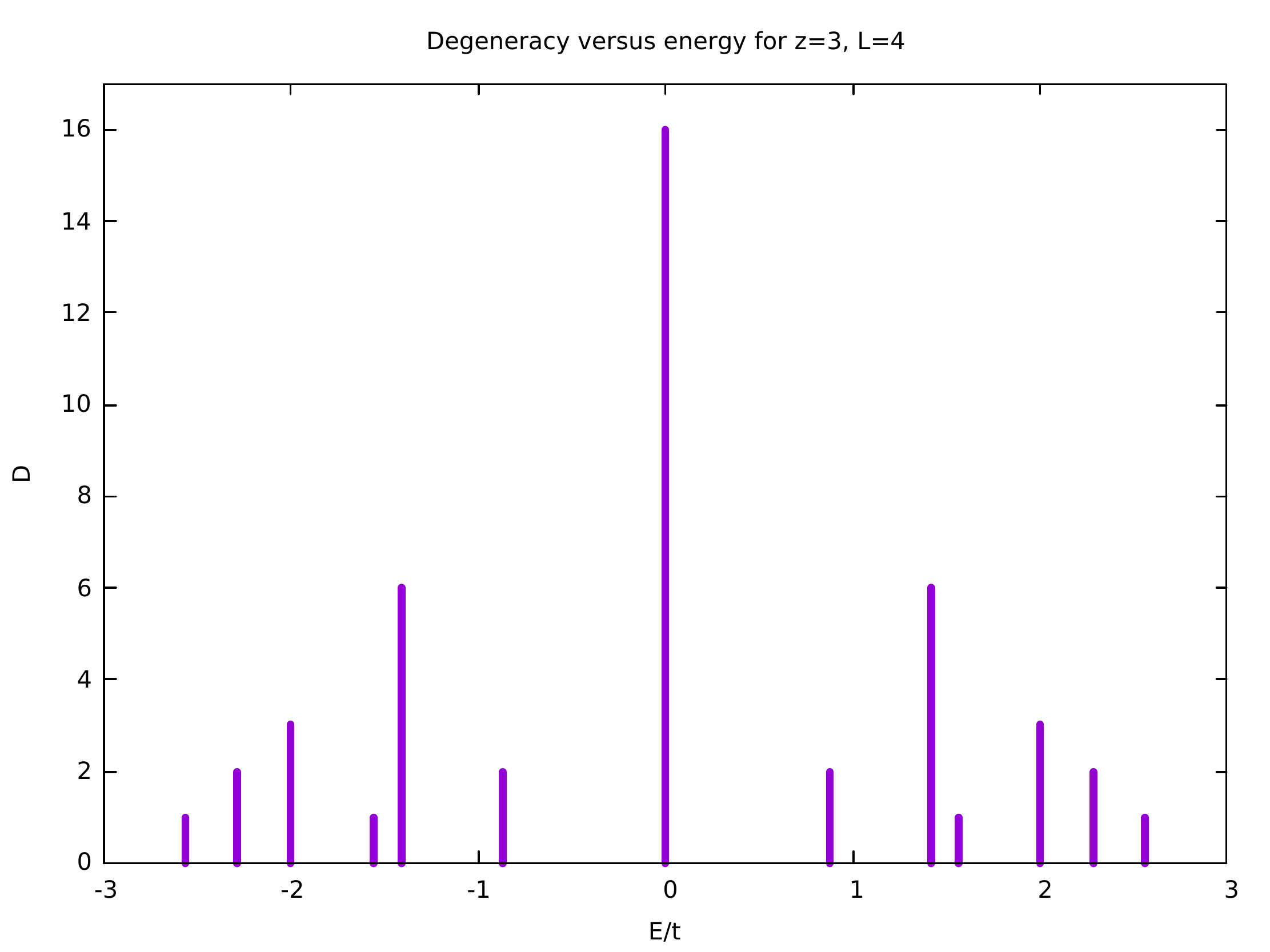}
  \caption{{Degeneracy of the energy levels of a Cayley tree of coordination number $z=3$ and $L=4$ shells}}.
  \label{L_4}
\end{figure}
\begin{figure}[h]
  \centering
  \includegraphics[width=0.5\columnwidth,clip]{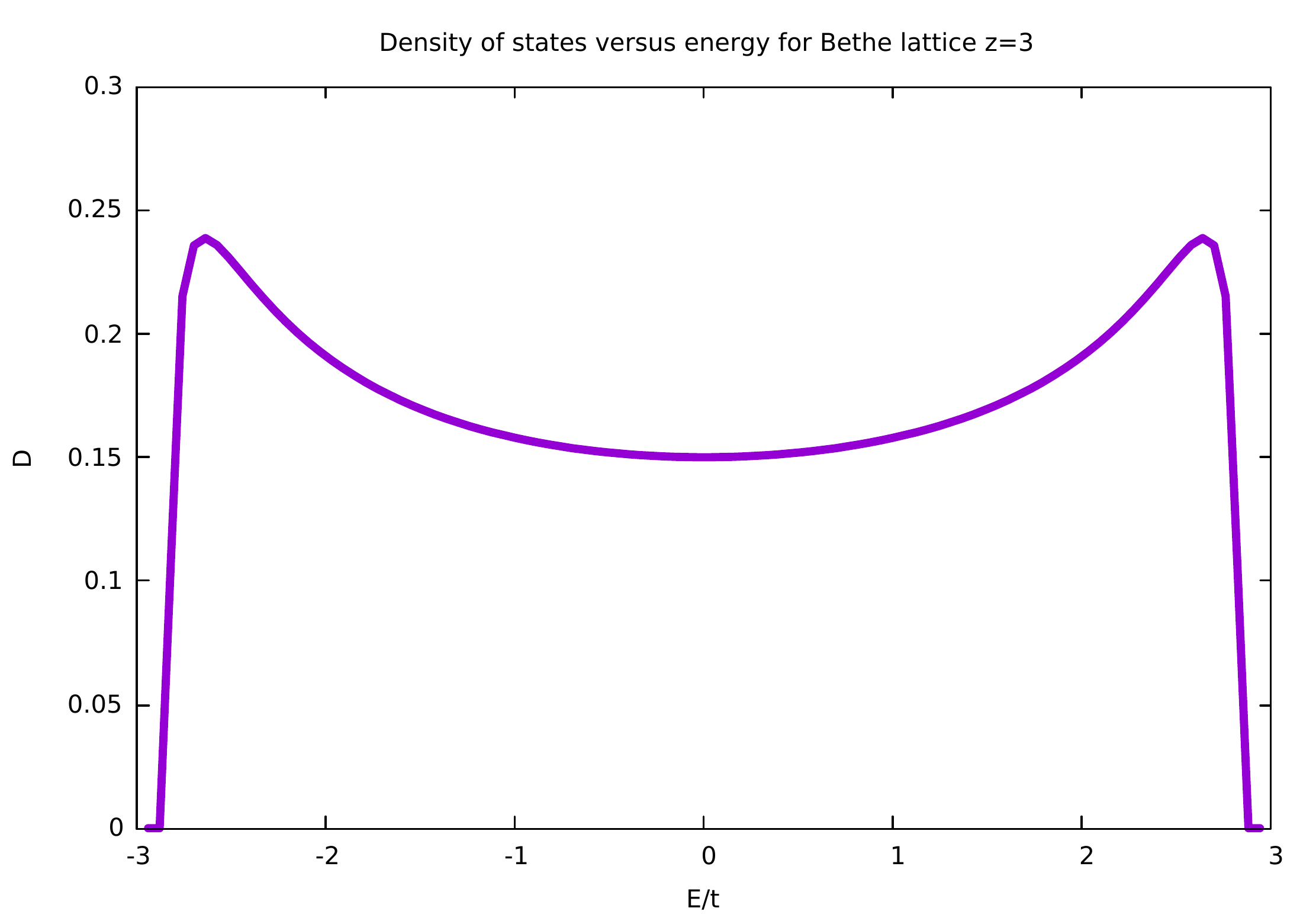}
  \caption{{Density of states of a Bethe lattice of coordination number $z=3$}}.
  \label{DOS_BL}
\end{figure}

  \begin{figure}[h]
  \centering
    \includegraphics[width=1.0\columnwidth,clip]{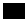}
  \caption{
    Spectrum of the symmetric eigenstates for a Cayley tree with {$z=3$ and $L=500$ shells}.
    The spectrum is obtained by solving the equation $\lambda P_L(\lambda)={z}P_{L-1}(\lambda)$ with the system (\ref{Recursive})
    via $10^7$ $x$ trials in the range
    $[-2\sqrt{2},2\sqrt{2}]$. White and black lines correspond to gap and solutions, respectively.
    {As the reader enlarges the picture,
    more detailed lines will appear.}
  }
  \label{fig2}
\end{figure}

\section{Discussion and conclusions}
We have analyzed the spectrum and the density of states of Cayley trees with coordination number $z$
and $L$ shells.  
First, note that our analysis is based on a simple recursive procedure that, starting from the boundary, \textit{i.e.}, the leafs of
the tree (in line with the classical case), resolves
the spectral problem by means of the canonical basis provided by the nodes.
{Second, our resulting 
  algorithm is extremely efficient and general as it can be applied to arbitrarily large trees of any degree $z$
  (see e.g. Fig. \ref{fig2}).}
Third, at no point have we relied on the ``apparent similarity'' between 
Bethe lattices and Cayley trees, not even in extrapolating our results in the thermodynamic limit $L\to\infty$.
Rather, as for the classical case~\cite{Ostilli},
we have demonstrated that, also in the present quantum case, 
the ``apparent similarity'' between Bethe lattices and Cayley trees is actually a quite misleading one,
as only the Cayley tree admits non symmetric eigenstates whose number grows exponentially as $(z-1)^L$,
a fact that in turn leads to a density of states dramatically different from that of
the Bethe lattice.

A careful analysis of the literature
on the subject shows that, despite their popularity, due to the above
``apparent similarity'', Bethe lattices and Cayley trees still lead to
misconceptions also in the quantum case.  Our results are all in
agreement with those reported in Ref.~\cite{Yorikawa} where, however,
the approach is essentially empirical and unclear from a theoretical
viewpoint, as well as in agreement with Ref.~\cite{Aryal} where,
however, the approach makes use of a special basis that applies for
$z=3$ and seems not extendable to $z$ even.  Our efficient recursive
approach lends itself to be applied to arbitrary (not necessarily
regular) trees and we believe that our work brings clarity to the
issue Cayley tree \textit{vs} Bethe lattice.

\section*{Acknowledgments}
We thank CNPq for funding (grants  302051/2018-0, 310561/2017-5 and 307622/2018-5).

\appendix
  \section{Illustrative examples with $z=3$}
  In this Section, we analyze in detail the spectra of Cayley trees with $z=3$ till $L=4$.
  Our aim is to illustrate with a concrete case the natural procedure that, starting from the leafs of the tree,
    produces all the spectrum showing also its structure and degeneracy.
 Indeed, one of the advantages of the canonical basis lies on the fact that
 there is no loss of generality in working out with a specific value of $z$.

 In what follows, we label the nodes as illustrated in Fig. \ref{fig1}
 {(for simplicity of notation, in the present case we replace the index of the first shall
1, 2 and 3, with the three letters $a$, $b$, and $c$, respectively)}.
 
 \subsection*{L=0}
 In this case $N=1$ and Eq. (\ref{psi}) trivially reduces to
\begin{eqnarray}
  \label{L0}
  0=E\psi(i=0)~,
\end{eqnarray}
which gives $E=0$.  

 \subsection*{L=1}
 In this case $N=4$ and Eqs. (\ref{psi}) give
 \begin{align}
     \label{L10}
  & E\psi(0)=t\left[\psi(a)+\psi(b)+\psi(c)\right]~,\\
    \label{L1a}
    & E\psi(a)=t\psi(0)~,\\
    \label{L1b}
    & E\psi(b)=t\psi(0)~,\\
    \label{L1c}
    & E\psi(c)=t\psi(0)~.
\end{align}
 We observe that, if $E\neq 0$, the three boundary components are all equal to
 \begin{align}
    \label{L1a1}
    & \psi(a)=\psi(b)=\psi(c)=\frac{t}{E}\psi(0)~,
\end{align}
 which, plugged into {Eq. (\ref{L1a}), give
 \begin{align}
     \label{L101}
  & E\psi(0)=\frac{3t^2}{E}\psi(0)~.
 \end{align}
On recalling that an eigenvector cannot coincide with the null vector, the last equation gives $E^2=3t^2$, i.e., $E=\pm \sqrt{3}t$. 
On the other hand, if we look for solutions with $E=0$, we see that these are satisfied if and only if 
 \begin{align}
     \label{L102}
  &\psi(a)+\psi(b)+\psi(c)=0~,\\
    \label{L1a2}
  &  \psi(0)=0,
\end{align}
which is a linear system with $\infty^2$ solutions, i.e., the eigenvalue $E=0$ is twofold degenerate.
The results are summarized in Table I.
\begin{table}[h!]
\begin{center}
{\caption{Energies and Degeneracies for $z=3$ and $L=1$}
  \begin{tabular}{ |c|c|} 
  \hline
   $E=\overset{~}{\sqrt{3}t}$ & $D=1$  \\
 $E=-\sqrt{3}t$ & $D=1$  \\ 
 $E=0$ & $D=2$  \\ 
 \hline
\end{tabular}}
\end{center}
\end{table}

 \subsection*{L=2}
 In this case $N=10$ and Eqs. (\ref{psi}) give
 \begin{align}
     \label{L20}
  & E\psi(0)=t\left[\psi(a)+\psi(b)+\psi(c)\right]~,\\
     \label{L21a}
     & E\psi(a)=t\left[\psi(0)+\psi(a1)+\psi(a2)\right]~,\\
     \label{L21b}
     & E\psi(b)=t\left[\psi(0)+\psi(b1)+\psi(b2)\right]~,\\
     \label{L21c}
     & E\psi(c)=t\left[\psi(0)+\psi(c1)+\psi(c2)\right]~,\\
     \label{L22}
     &E\psi(a1)=t\psi(a)~,\\
     &E\psi(a2)=t\psi(a)~,\\
     &E\psi(b1)=t\psi(b)~,\\
     &E\psi(b2)=t\psi(b)~,\\
     &E\psi(c1)=t\psi(c)~,\\
     &E\psi(c2)=t\psi(c)~.
\end{align}
 We observe that, if $E\neq 0$, the boundary components are 
 \begin{align}
    \label{L221}
    & \psi(a1)=\psi(a2)=\frac{t}{E}\psi(a)~,\\
    & \psi(b1)=\psi(b2)=\frac{t}{E}\psi(b)~,\\
    & \psi(c1)=\psi(c2)=\frac{t}{E}\psi(c)~,
\end{align}
 which, plugged into Eqs. (\ref{L21a})-(\ref{L21c}), give
 \begin{align}
     \label{L211}
     & \psi(a)=\psi(b)=\psi(c)=\frac{Et}{E^2-2t^2}\psi(0)~,
 \end{align}
 in turn, if $E^2-2t^2\neq 0$, plugging the last equation into Eq. (\ref{L20}) gives
 \begin{align}
     \label{L211}
     & E\psi(0)=\frac{3Et^2}{E^2-2t^2}\psi(0)~,
 \end{align}
 which provides $E^2=5t^2$. On the other hand, if we look for solutions with $E=0$, we see that these are satisfied if and only if 
 \begin{align}
     \label{L2110}
     & \psi(0)+\psi(a1)+\psi(a2)=0~,\\
     & \psi(0)+\psi(b1)+\psi(b2)=0~,\\
     & \psi(0)+\psi(c1)+\psi(c2)=0~,\\
     & \psi(a)=0~, \\
     & \psi(b)=0~, \\
     & \psi(c)=0~,
 \end{align}
 which is a linear system with $\infty^4$ solutions, i.e., the eigenvalue $E=0$ is fourfold degenerate.
 Similarly, if we look for solutions with $E^2-2t^2= 0$, we see that these are satisfied if and only if 
 \begin{align}
     \label{L2111}
     & \psi(a)+\psi(b)+\psi(c)=0~,\\
     & \psi(a1)=\psi(a2)=\frac{t}{E}\psi(a)~,\\
     & \psi(b1)=\psi(b2)=\frac{t}{E}\psi(b)~,\\
     & \psi(c1)=\psi(c2)=\frac{t}{E}\psi(c)~,
 \end{align}
which, for both $E=\sqrt{2}t$ and $E=-\sqrt{2}t$, is a linear system with $\infty^2$ solutions, i.e., the eigenvalues $E=\sqrt{2}t$ and $E=-\sqrt{2}t$ are twofold degenerate.
The results are summarized in Table II.
\begin{table}[h!]
\begin{center}
\caption{Energies and Degeneracies for $z=3$ and $L=2$}
\begin{tabular}{ |c|c|} 
 \hline
 $E=\overset{~}{\sqrt{5}t}$ & $D=1$  \\
 $E=-\sqrt{5}t$ & $D=1$  \\
 $E=\sqrt{2}t$ & $D=2$  \\
 $E=-\sqrt{2}t$ & $D=2$  \\ 
 $E=0$ & $D=4$  \\ 
 \hline
\end{tabular}
\end{center}
\end{table}

 \subsection*{L=3}
 In this case $N=22$ and Eqs. (\ref{psi}) give
 \begin{align}
     \label{L30}
  & E\psi(0)=t\left[\psi(a)+\psi(b)+\psi(c)\right]~,\\
     \label{L31}
     & E\psi(a)=t\left[\psi(0)+\psi(a1)+\psi(a2)\right]~,\\
     & E\psi(b)=t\left[\psi(0)+\psi(b1)+\psi(b2)\right]~,\\
     & E\psi(c)=t\left[\psi(0)+\psi(c1)+\psi(c2)\right]~,\\
     \label{L32}
     & E\psi(a1)=t\left[\psi(a)+\psi(a11)+\psi(a12)\right]~,\\
     & E\psi(a2)=t\left[\psi(a)+\psi(a21)+\psi(a22)\right]~,\\
     & E\psi(b1)=t\left[\psi(b)+\psi(b11)+\psi(b12)\right]~,\\
     & E\psi(b2)=t\left[\psi(b)+\psi(b21)+\psi(b22)\right]~,\\
     & E\psi(c1)=t\left[\psi(c)+\psi(c11)+\psi(c12)\right]~,\\
     & E\psi(c2)=t\left[\psi(c)+\psi(c21)+\psi(c22)\right]~,\\
     \label{L32}
     &E\psi(ad_1d_2)=t\psi(ad_1)~, \quad \forall~ d_1,d_2\in\{1,2\}~,\\
     &E\psi(bd_1d_2)=t\psi(bd_1)~, \quad \forall~ d_1,d_2\in\{1,2\}~,\\
     &E\psi(cd_1d_2)=t\psi(cd_1)~, \quad \forall~ d_1,d_2\in\{1,2\}~.
\end{align}
 We observe that, if $E\neq 0$, the boundary components are 
 \begin{align}
    \label{L321}
    & \psi(ad_1d_2)=\frac{t}{E}\psi(ad_1)~,\quad \forall~ d_1,d_2\in\{1,2\}~,\\
    & \psi(bd_1d_2)=\frac{t}{E}\psi(bd_1)~,\quad \forall~ d_1,d_2\in\{1,2\}~,\\
    & \psi(cd_1d_2)=\frac{t}{E}\psi(cd_1)~,\quad \forall~ d_1,d_2\in\{1,2\}~,
\end{align}
 which, plugged into Eqs. (\ref{L32}) give
 \begin{align}
    \label{L322}
    & (E^2-2t^2)\psi(ad_1)=Et\psi(a)~, \forall~ d_1 \in\{1,2\}~,\\
    & (E^2-2t^2)\psi(bd_1)=Et\psi(b)~, \forall~ d_1 \in\{1,2\}~,\\
    & (E^2-2t^2)\psi(cd_1)=Et\psi(c)~, \forall~ d_1 \in\{1,2\}~,
\end{align}
which, in turn, if $E^2-2t^2\neq 0$, give the system 
  \begin{align}
    \label{L323}
    & \psi(ad_1)=\frac{Et}{E^2-2t^2}\psi(a)~, \forall~ d_1 \in\{1,2\}~,\\
    & \psi(bd_1)=\frac{Et}{E^2-2t^2}\psi(b)~, \forall~ d_1 \in\{1,2\}~,\\
    & \psi(cd_1)=\frac{Et}{E^2-2t^2}\psi(c)~, \forall~ d_1 \in\{1,2\}~,\\
    & \psi(a)=\psi(b)=\psi(c)=\frac{(E^2-2t^2)t}{E(E^2-2t^2)-2Et^2}\psi(0),
  \end{align}
 which, in turn, if $E(E^2-2t^2)-2Et^2\neq 0$, provides finally the equation
 \begin{align}
    \label{L324}
    & E\psi(0)=\frac{3t^2(E^2-2t^2)}{E(E^2-2t^2)-2Et^2}\psi(0)~,
\end{align}
 i.e.,
 \begin{align}
   \label{L324}
   E^4-7E^2t^2+6t^4=0~,
\end{align}
which is solved for $E=\pm t$, and $E=\pm \sqrt{6}t$.  
On the other hand, if we look for solutions with $E=0$, we see that these are satisfied if only if 
 \begin{align}
     \label{L301}
     & \psi(a)+\psi(b)+\psi(c)=0~,\\
     & \psi(0)=0~,\\
     &\psi(ad_1)=0, \quad \forall~ d_1 \in\{1,2\}~,\\
     &\psi(bd_1)=0, \quad \forall~ d_1 \in\{1,2\}~,\\
     &\psi(cd_1)=0, \quad \forall~ d_1 \in\{1,2\}~,\\
     & \psi(a)+\psi(a11)+\psi(a12)=0~,\\
     & \psi(a)+\psi(a21)+\psi(a22)=0~,\\
     & \psi(b)+\psi(b11)+\psi(b12)=0~,\\
     & \psi(b)+\psi(b21)+\psi(b22)=0~,\\
     & \psi(c)+\psi(c11)+\psi(c12)=0~,\\
     & \psi(c)+\psi(c21)+\psi(c22)=0~,
\end{align}
which is a linear system with $\infty^8$ solutions, i.e., the eigenvalue $E=0$ is eightfold degenerate.
If instead we look for solutions with $E^2-2t^2=0$, from Eqs. (\ref{L322}) we see that these are satisfied if only if
 \begin{align}
     \label{L302}
     & \psi(0)=\psi(a)=\psi(b)=\psi(c)=0~,\\
     & \psi(a1)+\psi(a2)=0~,\\
     & \psi(b1)+\psi(b2)=0~,\\
     & \psi(c1)+\psi(c2)=0~,
 \end{align}
which is a linear system with $\infty^3$ solutions, i.e., each of the eigenvalues $E=\pm \sqrt{2}t$ is threefold degenerate.
Finally, if we look for solutions with $(E^2-2t^2)-2t^2=0$, we see that these are satisfied if only if
 \begin{align}
     \label{L302}
     & \psi(a)+\psi(b)+\psi(c)=0~,
 \end{align}
which is a linear system with $\infty^2$ solutions, i.e., each of the eigenvalues $E=\pm 2 t$ is twofold degenerate.
The results are summarized in Table III.
\begin{table}[h!]
\begin{center}
\caption{Energies and Degeneracies for $z=3$ and $L=3$}
\begin{tabular}{ |c|c|} 
  \hline
  $E=t$ & $D=1$  \\
  $E=-t$ & $D=1$  \\  
 $E=\sqrt{6}t$ & $D=1$  \\
 $E=-\sqrt{6}t$ & $D=1$  \\
  $E=0$ & $D=8$  \\
  $E=\sqrt{2}t$ & $D=3$  \\
 $E=-\sqrt{2}t$ & $D=3$  \\ 
  $E=2t$ & $D=2$  \\
 $E=-2t$ & $D=2$  \\ 
  \hline
\end{tabular}
\end{center}
\end{table}

 \subsection*{L=4}
 In this case $N=46$ and Eqs. (\ref{psi}) give the results
summarized in Table IV.
\begin{table}[h!]
\begin{center}
\caption{Energies and Degeneracies for $z=3$ and $L=4$}
\begin{tabular}{ |c|c|} 
  \hline
%
  $E=\frac{1+\sqrt{17}}{2}t$ & $D=1$  \\
  $E=-\frac{1+\sqrt{17}}{2}t$ & $D=1$  \\
  $E=\frac{1-\sqrt{17}}{2}t$ & $D=1$  \\
  $E=-\frac{1-\sqrt{17}}{2}t$ & $D=1$  \\
  $E=\sqrt{3+\sqrt{5}}t$ & $D=2$  \\
  $E=-\sqrt{3+\sqrt{5}}t$ & $D=2$  \\
  $E=\sqrt{3-\sqrt{5}}t$ & $D=2$  \\
  $E=-\sqrt{3-\sqrt{5}}t$ & $D=2$  \\
  $E=0$ & $D=16$  \\
  $E=2t$ & $D=3$  \\
  $E=-2t$ & $D=3$  \\
  $E=\sqrt{2}t$ & $D=6$  \\
  $E=-\sqrt{2}t$ & $D=6$  \\
  \hline
\end{tabular}
\end{center}
\end{table}

\end{document}